# Variable Emissivity Modeling for Sustainable Lunar Surface Habitats

Sam Keller[1] and Ognjen Ilic[2]
*University of Minnesota – Twin Cities, Minneapolis, Minnesota, 55455*

Sarah Stewart[3] and Sydney Taylor[4]
*NASA Johnson Space Center, Houston, Texas, 77058,*

**Lunar habitats will be one of the first platforms to enable long-term human presence beyond Low Earth Orbit. These structures act as a stepping stone for exploring our solar system while simultaneously enabling lunar resource utilization, low-energy cryopreservation, and various other applications. These habitats must be designed to withstand the extreme thermal variation of the lunar surface caused by the changing orientation with respect to the Sun and Earth. White paints and multi-layered insulation are conventionally used to minimize solar heating, yet this approach is static and results in a structure that requires internal heating to survive lunar night. An adaptive approach to control absorbed and emitted radiation allows for highly efficient daytime cooling and improved nighttime heat retention. Louvers and shutters have been employed to switch between high- and low-emissivity states; however, this approach relies on ensuring moving parts are resilient to dust contamination. Alternatively, variable emissivity materials are a solid-state solution with no moving parts. The emissivity of these materials can be switched passively based on surface temperature, or actively as a result of applied voltage. Despite their potential to reduce power consumption and increase thermal stability, variable emissivity materials have yet to be explored on the lunar surface. We first present a finite element modeling approach to predict the thermal performance of simplified habitats in realistic lunar environments. We then demonstrate the benefit of variable emissivity materials for thermal stability and lunar night survival by comparing them to traditional constant emissivity coatings. By using variable emissivity materials, we envision near-constant temperature lunar habitats with significantly reduced internal heating requirements.**

## Acronyms and Nomenclature

CM = COMSOL Multiphysics®
JPL = Jet Propulsion Lab
LEO = low earth orbit
LRO = lunar reconnaissance orbiter
LSH = lunar surface habitat
NASA = National Aeronautics and Space Administration
PLSS = portable life support systems
TD = Ansys Thermal Desktop®
VEM = variable emissivity material
$\alpha$ = solar absorptivity
$c_p$ = specific heat
$\varepsilon$ = thermal emissivity
$h$ = heat transfer coefficient

---

[1] NSTGRO Fellow, Department of Mechanical Engineering, 111 Church Street SE, Minneapolis, MN 55455.
[2] Assistant Professor, Department of Mechanical Engineering, 111 Church Street SE, Minneapolis, MN 55455.
[3] Thermal Analyst, Thermal Design Branch, NASA JSC, ES3
[4] Orion PTCS Analysis Lead, Thermal Design Branch, NASA JSC, ES3

| | | |
|---|---|---|
| $k$ | = | thermal conductivity |
| $\rho$ | = | density |
| $\phi$ | = | solar diameter angle |
| $Q_{solar}$ | = | solar heat flux |
| $t$ | = | thickness |
| $T$ | = | temperature |
| $\theta$ | = | solar elevation angle |

## I. Introduction

Establishing a sustainable and long-term human presence on the moon is a critical step towards lunar resource utilization and making the first footprints on Mars. To achieve this vision, NASA plans to construct a moon base by the 2030s, which will be one of the many key elements of the agency's lunar surface architecture subject to the dynamic thermal conditions of the lunar south pole. Habitats, landing vehicles, and other critical structures must be designed to survive the highly contrasting radiative environments of day and night on the lunar surface. Due to the lack of an atmosphere, solar radiation can cause extreme daytime heating, which often requires externally facing radiators to balance absorbed, emitted, and generated heat. These radiators are designed to reflect solar radiation in the ultraviolet to visible spectrum and emit thermal radiation across near-infrared wavelengths [1]. Unless properly acutated or stored, these radiators will become a critical point of heat loss during lunar night. Recent studies have demonstrated how folding these radiators can be used to reduce their surface area, in turn minimizing heat loss for lunar night survival [2]. Despite significantly reducing heat loss, this approach relies on mechanical actuation, which is susceptible to cyclic failure and increases the overall mass and power consumption of the radiator sub-system. Additionally, the presence of lunar dust results in a significant failure risk for mechanical actuators.

Recent studies have identified variable emissivity materials (VEMs) as a suitable approach to dynamic radiative thermal management in systems ranging from satellites operating in Low Earth Orbit (LEO) to space suits and Portable Life Support Systems (PLSS) [3]–[6]. VEMs directly control heat rejection via a solid-state phase transition to achieve high and low thermal emissivity states without the need for mechanical actuation [7]. Thermochromic VEMs are materials which change emissivity as a function of their temperature in a reversible manner. Of these materials, Vanadium Dioxide ($VO_2$) is the most commonly studied for thermal management in space. $VO_2$ displays a reversible insulator-to-metal phase change at 68°C; however, this can be tuned to lower temperatures through metal doping [8]. The phase change reorganizes the material at an atomic level, changing its dielectric properties and creating stable states with high and low thermal emissivity. Recent studies using $VO_2$ have demonstrated exceptional emissivity control, reaching as high as 0.92 in the high state and 0.095 in the low state [9]. Thermochromic VEMs are often considered a passive thermal management solution, requiring net-zero power to switch between heating and cooling [10]. Alternatively, electrochromic materials and chalcogenide glasses are examples of active variable emissivity materials, which require applied voltage and joule heating, respectively, to induce a phase change [11], [12].

Despite many studies designing VEMs for generalized use on Earth and in space, understanding the benefit on realistic systems remains a key shortfall preventing widespread adoption. Recent demonstrations have shown how these materials can be modeled in LEO using *Thermal Desktop®* [13]. We expand upon this method by adapting modeling practices to the lunar surface. Additionally, we approach this problem independently of specific materials, noting that the method provided can easily be adapted to both thermochromic (passive) and electrochromic (active) variable emissivity materials. We first outline the procedure necessary to accurately model the radiative environment of a real location on the lunar surface. We then develop a representative lunar habitat with body-mounted radiators similar to the concept highlighted in Figure 1 and use this to demonstrate how VEMs can be used to reduce heat loss during lunar night. Notably, we use both *Thermal Desktop®* and *COMSOL Multiphysics®* to demonstrate the unique solutions each program requires to vary surface emissivity in a transient simulation.

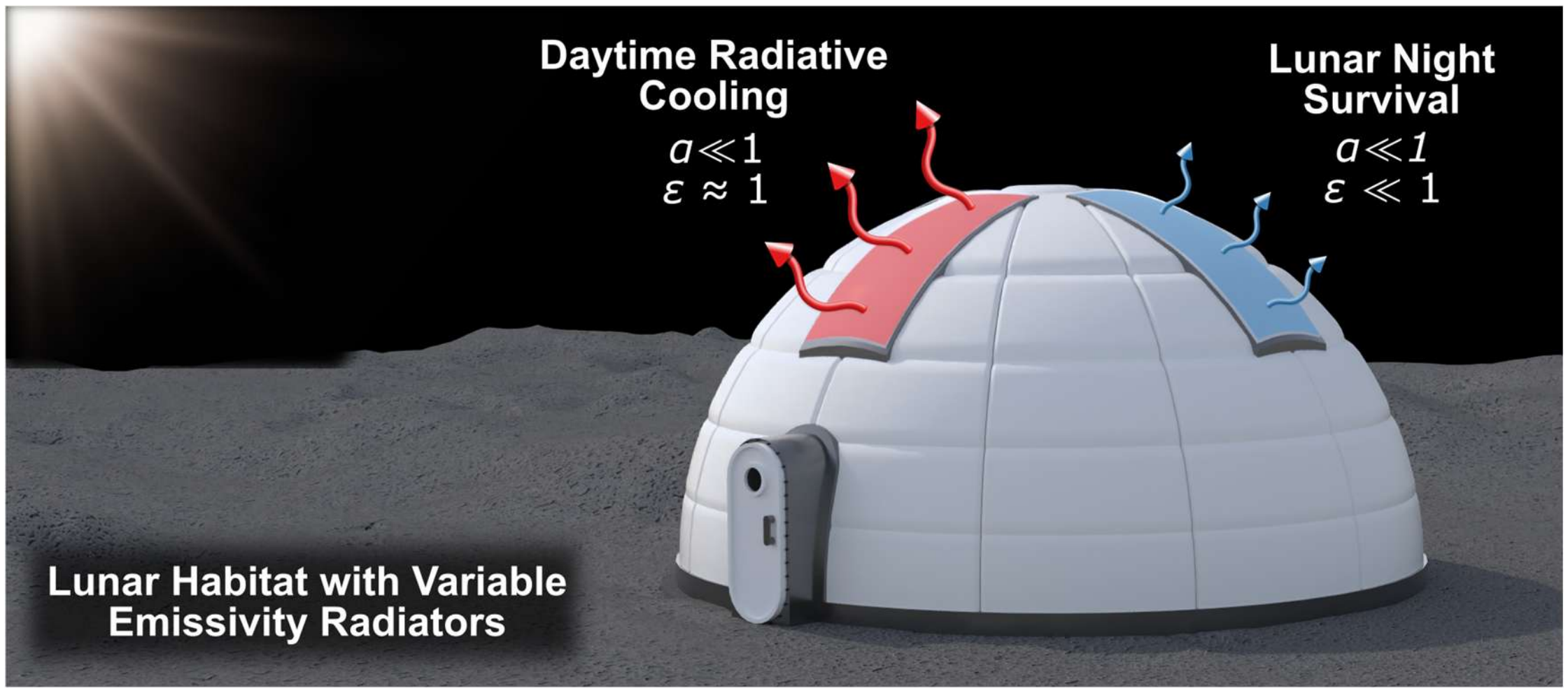

**Figure 1. Daytime radiative cooling and nighttime heat retention via variable emissivity radiators designed for sustainable lunar habitats.**

## II. Methods

### A. Lunar Surface Thermal Model

To demonstrate the benefits of variable emissivity materials on the lunar surface, we must account for all forms of radiative heat transfer. This includes solar irradiation, earthshine, and emitted and reflected radiation from the lunar surface. Additionally, view factors to the lunar surface and regional topography can significantly impact radiative heat transfer at the lunar poles, where the solar elevation angle remains low year-round. Many studies have detailed various thermal models for different regions on the moon [14], [15]. As such, we approach this complex environment from the ground up by first assembling a two-layer planar model for the lunar surface and validating surface temperatures against known data collected during Apollo 17 [16]. We expand upon this model by implementing elevation data collected by the Lunar Reconnaissance Orbiter (LRO) to properly account for regional shading and slope [17]. We then return to a planar geometry, which is correlated with the topographic model to accurately describe the radiative heat exchange at a known location while reducing the complexity and computation time associated with the topographic model. This correlated planar thermal model is then used for all subsequent studies of lunar habitats and variable emissivity radiators.

We begin by establishing a planar two-layer model consisting of an upper "fluff" layer and a lower "regolith" layer to accurately model the lunar surface temperature. This model, along with preliminary studies to produce a simplified, realistic lunar environment, was entirely conducted using *COMSOL Multiphysics*®. Subsequent studies of variable emissivity materials were approached through both *COMSOL Multiphysics*® and *Ansys Thermal Desktop*® simultaneously to highlight their unique implementation procedures. A two-layer lunar surface model is used throughout this study to more accurately model surface temperature associated with a low-density top layer. It should be noted that the through-thickness temperature distribution was not considered throughout this study, as we assume no conduction from the habitat to the lunar surface. The lunar surface model, along with all associated thermo-optical properties, can be found in Table 1. This model was derived from previous studies examining cryogenic fluid storage at the lunar South Pole [18].

**Table 1. Thermal and optical properties of the two-layer lunar surface model [18].**

| Property | Fluff | Regolith |
|---|---|---|
| Thickness, m | 0.02 | 0.60 |
| Thermal Conductivity, W/(mK) | $9.22*10^{-4}(1+1.48(T/350)^3)$ | $9.30*10^{-3}(1+0.73(T/350)^3)$ |
| Density, kg/m$^3$ | 1000 | 2000 |
| Specific Heat, J/(kgK) | 1050 | 1050 |
| Thermal Emissivity | 0.97 | - |
| Solar Absorptivity | 0.87 | - |
| Internal Heat Flow, W/m$^2$ | - | 0.031 |

We use a solar absorptivity of 0.87 and a thermal emissivity of 0.97 to represent a hot-case dynamic environment. A more accurate model may be created by updating these values in future studies to better match specific regions on the moon. A schematic of this model with all incoming and outgoing heat sources is found in Figure 2a). Solar elevation and azimuthal angles were produced using the JPL Horizons System and used to compare the modeled surface temperature to data collected during Apollo 17. This comparison is found in Figure 2b), where we see a strong correlation between the model and experimental measurements. This confirmed an accurate representation of the lunar surface and allowed us to continue with implementing the effects of regional topography and view factor for a selected location.

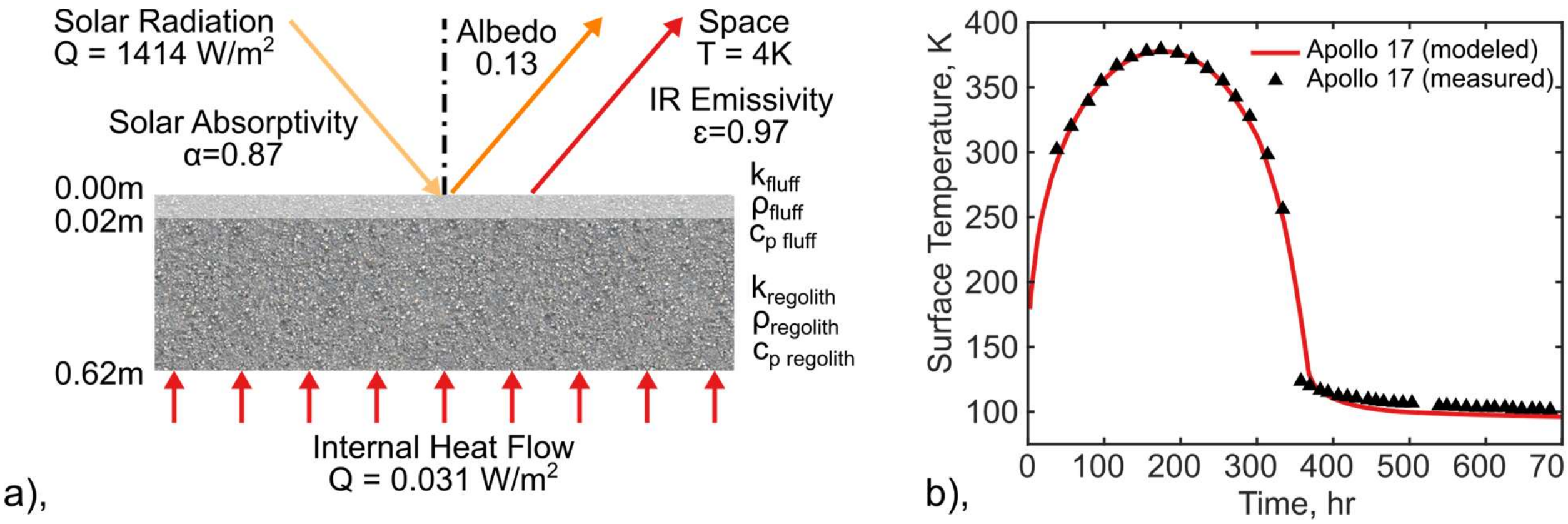

**Figure 2. a), Two-layer thermal model used to describe the lunar surface. b), Comparison of surface temperature using this model to temperature probe data collected during Apollo 17.**

The lunar south pole is the primary area of focus for this study due to the interest in in-situ resource utilization, cryopreservation, and current planned missions to establish a long-term presence on the moon [19], [20]. It should be noted that the benefit of variable emissivity radiators may be greater as one approaches more dynamic environments near the equator. To narrow the study further, we select the de Gerlache Rim 2 site (88.161°S, 305.936°E) as the location for case studies conducted throughout this work. This selection was made from the candidate landing sites for the Artemis missions due to its proximity to the south pole [21]. The topography of this region, along with a centralized probe to track surface temperature, can be seen in Figure 3a). This point will be used to correlate a simplified planar model to further reduce computation time when evaluating a representative lunar habitat. We pursue initial topographic modeling using *COMSOL Multiphysics*® to greatly reduce the amount of pre-processing required to convert a digital elevation map to a two-layer thermal model. We achieve this conversion by importing the digital elevation map as an image function and mapping this to a parametric surface, which is scaled according to the maximum and minimum latitude, longitude, and elevation. Once the parametric surface is assembled, it is extruded to create two domains representative of the regolith and fluff. Each domain is then meshed with 60m horizontal accuracy to match the effective resolution of the digital elevation map. This meshed domain is presented in Figure 3b).

In studies where a habitat is included, it is no longer computationally efficient or practical to resolve temperatures for the entire domain due to differences in heat transfer length scales. As such, we assemble two separate planar models shown in Figure 3c). These planar models have identical through-thickness discretization to the topographic model; however, the surface area is only 100m in diameter. We first examine a horizontal planar model and then update the rotation to match the local slope measured at the same point as our temperature probe presented in Figure 3a). This study will demonstrate the importance of including both slope and topographic features when calculating lunar surface temperature near the south pole. Our simplified lunar habitat design will be placed on the planar surface with local slope correction to accurately predict energy savings and lunar night survivability in a realistic location on the moon.

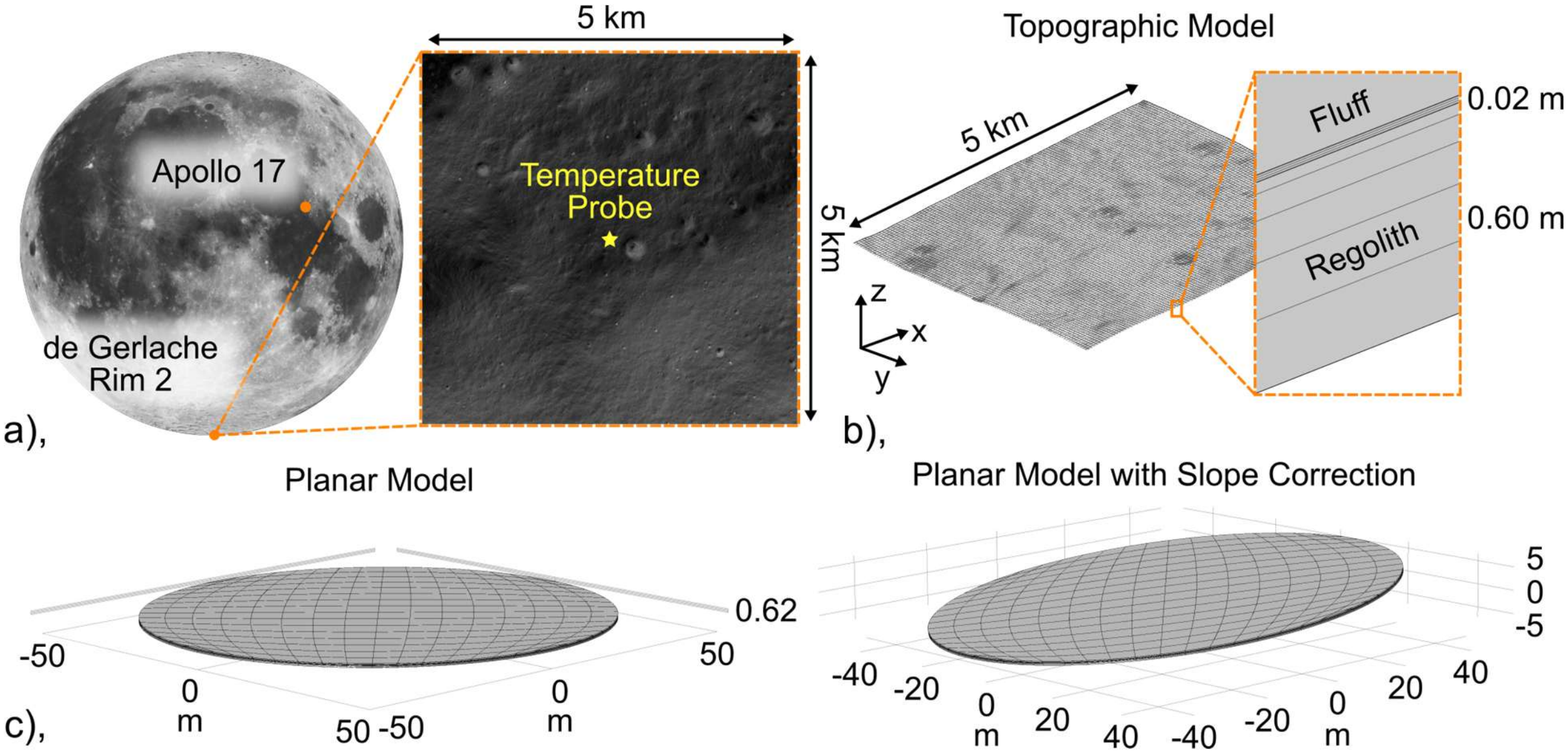

**Figure 3. a), Topographic map of the de Gerlache Rim 2 site near the lunar south pole (88.161°S, 305.936°E) with a centralized surface temperature probe. b), Visualization of the two-layer mesh used in COMSOL Multiphysics®. c), Simplified planar models with and without slope correction.**

With the region of interest selected, we then choose an appropriate time period to model surface temperature by evaluating the solar elevation angle at this location. The oscillating elevation angle over one full year can be found in Figure 4, where we identify a two day-night cycle period, which will be used to define the start and stop times of our transient simulation. We select a time frame with the greatest solar elevation angles, as this will result in the most thermally dynamic environment, clearly demonstrating the application for variable emissivity materials. Additionally, we select two full day-night cycles to ensure results are minimally influenced by the initial conditions.

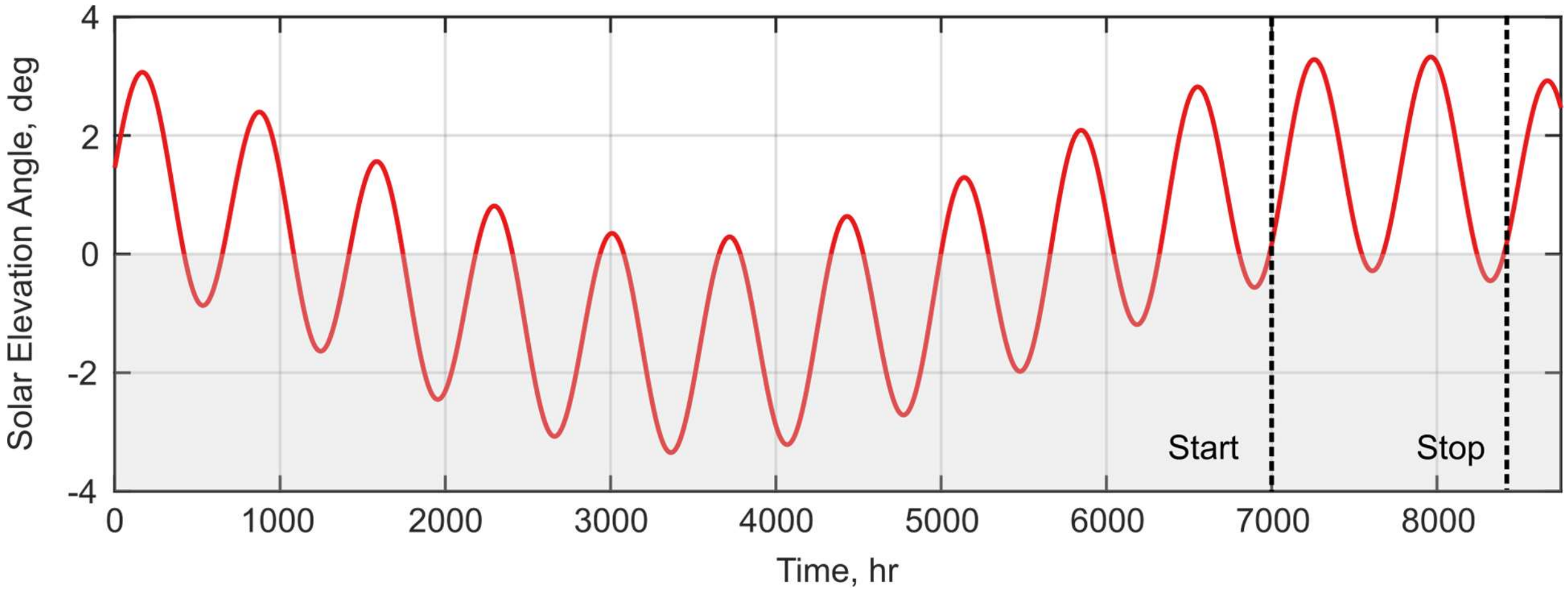

**Figure 4. Solar elevation angle observed at the de Gerlache Rim 2 site (88.161°S, 305.936°E) over one full year. Start and stop times indicated here are used throughout this work.**

Combining the transient solar vectors with our three-dimensional lunar surface model provides a detailed simulation that accounts for shading associated with local topography; however, it is equally critical to account for solar occultation with respect to the horizon. At either of the lunar poles, the solar elevation angle, $\theta$, is comparable to the solar diameter angle, $\phi$, resulting in a fraction of the visible sun appearing below the horizon for significant periods of the day. As such, the solar heat flux should be proportional to the fraction of the sun visible above the horizon. We approach this by attenuating heat flux using the following equation, outlined in NASA HLS-UG-001 [22]:

$$Q_{solar} = 1414\ \left[\frac{W}{m^2}\right] * \left(1 - \frac{\operatorname{acos}\left(\frac{tan\theta}{tan\phi}\right) - \frac{tan\theta}{tan\phi}\sqrt{1-\left(\frac{tan\theta}{tan\phi}\right)^2}}{\pi}\right) \text{ for } -\phi \le \theta \le \phi \tag{1}$$

Incorporating solar occultation into our model ensures we are accurately representing the radiative environment by accounting for regional attenuation when the sun partially crosses the horizon. We run a transient simulation of the de Gerlache Rim 2 site, where the surface temperatures for the entire region over the second day-night cycle can be observed in Figure 5a). The temperature fluctuations over two day-night cycles measured at the central probe on the topographic model are plotted in Figure 5b) alongside the average surface temperatures of our planar models with and without slope and shading correction. Shading correction for the planar model is achieved by manually attenuating the solar heat flux to match the topographic model. This is done by recording the times at which solar irradiation is zero on the topographic model and setting solar heat flux to zero over these time frames in the planar model. The effects of this correction are primarily observed at the onset of lunar night. When comparing the two planar models, we see drastically different temperature profiles, which we attribute to a greater maximum solar incidence angle during the day for the slope-matched model. Additionally, we see a strong correlation between the simplified planar model with shading and slope correction and the probed location on the topographic model. This indicates that our approach to drastically simplifying the model complexity continues to accurately represent the radiative environment. Notably, there is a ~6K decrease in surface temperature for the planar model during lunar night. This is attributed to a greater view factor to space when compared to the probed location of the topographic model, allowing for marginally increased nighttime heat rejection. The modeling approach presented provides us with a simplified platform that accounts for local and regional shading to produce a highly realistic representation of the radiative environment on the lunar surface.

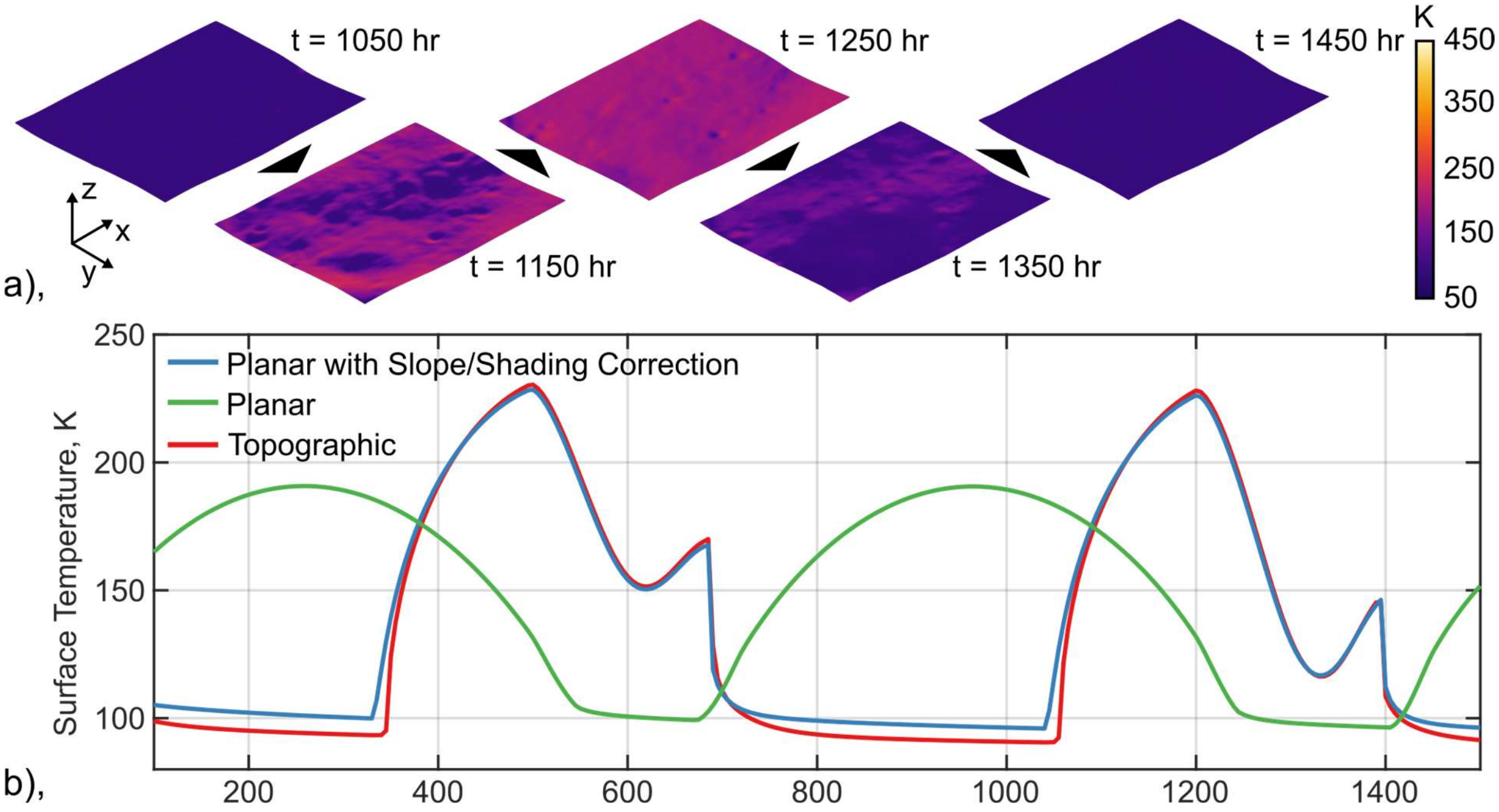

**Figure 5. a), Surface temperatures of the de Gerlache Rim 2 site for one full day cycle. b), Surface temperatures of the centralized probe region in comparison to a horizontal planar model and a planar model where the slope and shading have been matched to the topographic model.**

### B. Lunar Habitat Thermal Model

A representative thermal model for a lunar habitat must first be developed, which will then serve as a platform for investigating variable emissivity materials. Current habitat concepts often exist in the general form of cylinders, toroids, or hemispheres with a largely inflatable volume [23]. We simplify these concepts into a single dome structure with representative thermal properties for the habitat wall and interior environment. We then mount externally facing radiators on the body of the habitat, noting that they are connected to the interior wall via a high-conductivity pathway to simulate a realistic radiator implementation. In particular, we size the habitable volume and external radiators to match previous case studies on NASA's conceptual Lunar Surface Habitat [2]. It is assumed that the outermost layer of the inflatable habitat is made of beta cloth, resulting in a solar absorptivity of approximately 0.4 and thermal emissivity of 0.9 [24]. The habitat design and relevant thermo-optical properties can be seen in Figure 6a). These properties represent and approximate an average of the materials used in NASA's TransHab design for an inflatable habitat on the International Space Station [25]. Lastly, we assume the habitat is conductively insulated from the lunar surface. We acknowledge that this design represents a simplified habitat configuration, providing us with a platform to compare variable emissivity radiators to their conventional static counterparts. We assume the variable emissivity radiators exhibit a solar absorptivity of 0.2 and thermal emissivity modulation between 0.2 and 0.8. These values represent common target metrics when designing variable emissivity materials, which allows us to investigate VEMs without assuming the use of specific structures or materials.

This habitat model, coupled with the two-layer planar surface, was independently assembled in *COMSOL Multiphysics®* and in *Ansys Thermal Desktop®*. The use of variable emissivity radiators requires unique modeling approaches for both programs. In particular, a switch must be implemented to pause the simulation once selected criteria are met, change the emissivity of the surface, and re-calculate the radiosity at each time step. In *COMSOL Multiphysics®*, we use the *Events* interface to establish emissivity modulation conditions, while in *Ansys Thermal Desktop®,* we achieve a similar goal through the use of *Dynamic Sinda* [26]. We outline this general procedure in Figure 6b), where we provide two emissivity modulation criteria that will be demonstrated in this work. We first implement an active switch to match the onset of day and night, demonstrating how VEMs could be used to support lunar night survival by significantly reducing heat rejection. A passive approach will then be presented where we apply a survival heater to the habitat wall and couple its control with radiator emissivity to maintain near-constant temperatures over a full day-night cycle.

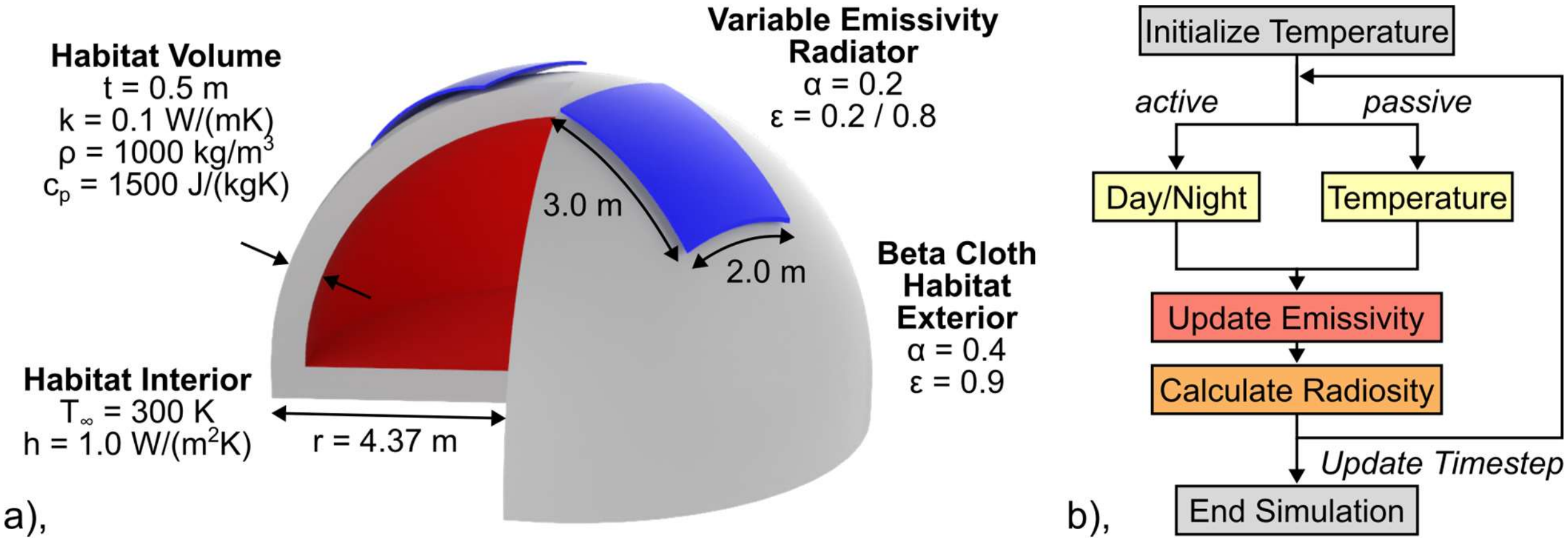

**Figure 6. a), Simplified lunar habitat design with body-mounted radiators. b), Modeling flowchart for actively and passively controlling variable emissivity materials.**

## III. Results

We first investigate the effect of actively modulating variable emissivity materials in response to the onset of lunar day and night. This study aims to present variable emissivity materials as an effective approach to reduce power consumption for lunar night survival. The simplified habitat is placed on a planar surface model representing the de Gerlache Rim 2 site, and the transient model is run for two day-night cycles. The habitat consists of two 2m x 3m radiators as shown in Figure 6a), which are connected to the interior wall via a high conductivity pathway and the interior temperature is set as a constant. We compare the average heat rejection per radiator for two unique cases varying radiator optical properties. Initially, the thermal emissivity of each radiator is set as a constant 0.8, arbitrarily representing radiator coatings such as white paints and optical solar reflectors. Then, a variable emissivity material is simulated by setting thermal emissivity to 0.2 at night to increase heat retention and 0.8 during the day to minimize overheating when in view of the sun. We assume the emissivity transition to be stepped, though if a specific VEM is known, this may be replaced by emissivity as a function of temperature over the phase transition range. The thermal emissivity of the habitat and radiators in each emissivity state can be viewed in Figure 7a). The heat rejection for each case study is reported in Figure 7b). Slight variation between *CM* and *TD* results can be attributed to different ray tracing algorithms between programs and numerical error associated with extremely low elevation angles at the beginning and end of the day. On average, we observe a 1.34 kW/m$^2$ decrease in heat flux between VEM and high emissivity case studies using *CM* and 1.38 kW/m$^2$ using *TD*. For a single lunar night from 685 hr to 1040 hr, we report a total energy savings of 268 kWh in *CM* and 275 kWh in *TD* for a 6 m$^2$ radiator.

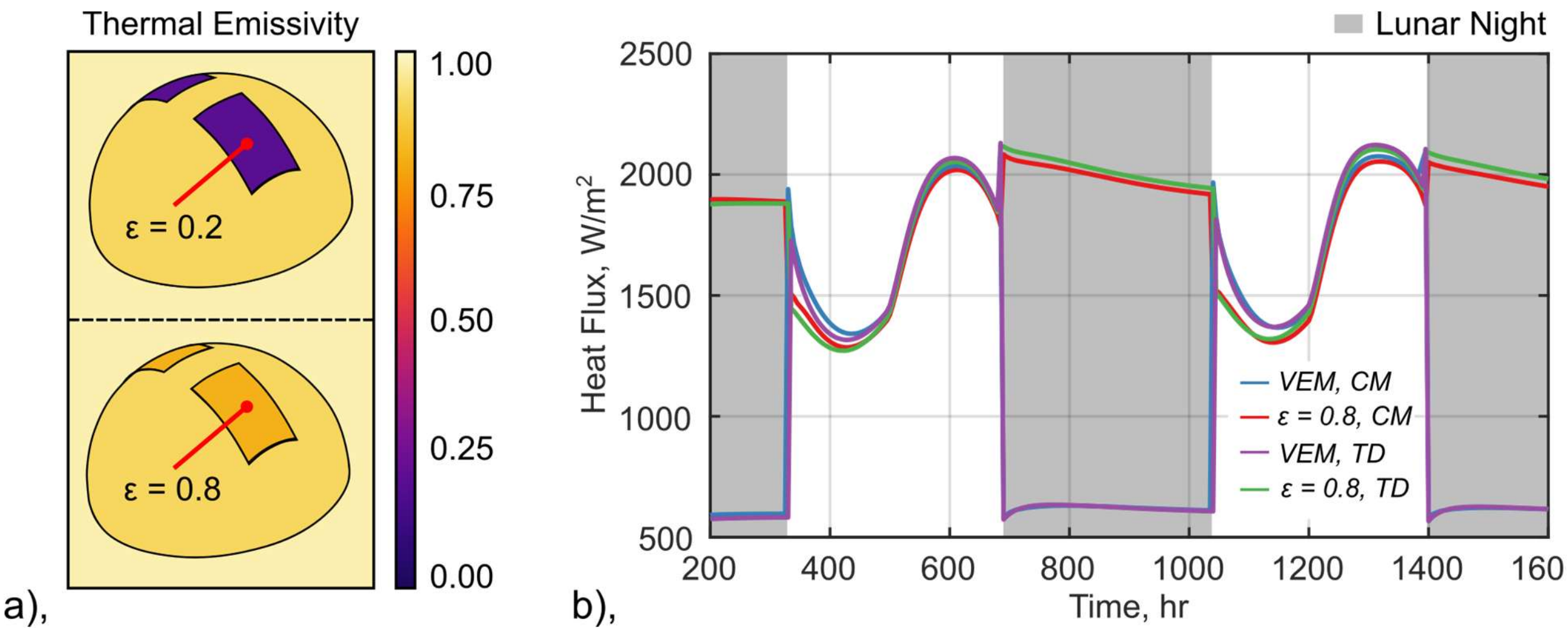

**Figure 7. a), Thermal emissivity of the simplified habitat and surrounding lunar surface in low and high emissivity states. b), Heat rejection of a 2m x 3m variable emissivity radiator and constant, high-emissivity radiator over two day/night cycles.**

This study presents a simplified control scheme for variable emissivity materials, where it is desirable to retain as much heat as possible during night and be highly emissive during the daytime. It should be noted that this form of control, while reducing power consumption at night, may not be the most efficient for daytime thermal stability. In a realistic system with variable emissivity radiators, internal heat generation should be coupled via closed-loop control to the radiator state. In doing so, one could further reduce power consumption during the day by limiting how much heat is rejected in response to varying internal and external loads. In subsequent case studies, we show variable emissivity radiators can be used to achieve near-constant internal temperatures without overly complicating the simplified lunar habitat model. To demonstrate thermal stability, we apply a heating element to the inner surface of the habitat wall. This heater operated in binary states, either functioning in the "on" state and supplying 5 kW to the wall when the temperature is below 290 K or turning off once the wall temperature reaches 295 K. In *Ansys Thermal Desktop®*, this was achieved using a boundary heating source and specifying on/off temperatures; however, in *COMSOL Multiphsyics®* this heater must be controlled using the *Events* interface, similarly to variable emissivity materials.

In addition to including a multi-state heater within the habitat, one must also change how the variable emissivity radiators switch state. Rather than modulating between high and low emissivity in response to day/night cycles, the state should now depend on the temperature of interest within the habitat. In our case, we probe the internal wall temperature to ensure it remains near room temperature, though this temperature measurement could easily be placed on the radiator itself to more closely represent the passive control of thermochromic VEMs such as vanadium dioxide. A set-point temperature of 292.5 K is defined, where above this the radiators operate in a high emissivity state, and below this they have low emissivity. Combining emissivity and heater control allows us to maintain increased thermal stability, notably by decreasing the time it takes to heat or cool to the set-point temperature. As shown in Figure 8), using a VEM allows us to achieve stable temperatures over lunar day and night, whereas replacing the variable emissivity radiator with a constant high emissivity radiator results in significantly lower temperatures at night. Additionally, a 5 kW heater does not provide sufficient heat for the simplified habitat presented in this work to operate within the maximum and minimum temperature thresholds previously specified for the heater. As such, the high-emissivity case study requires heaters to be constantly powered. While this result is highly dependent on the geometry and thermal properties of the habitat, it demonstrates how VEMs can be used to reach more habitable temperatures without increasing heat load.

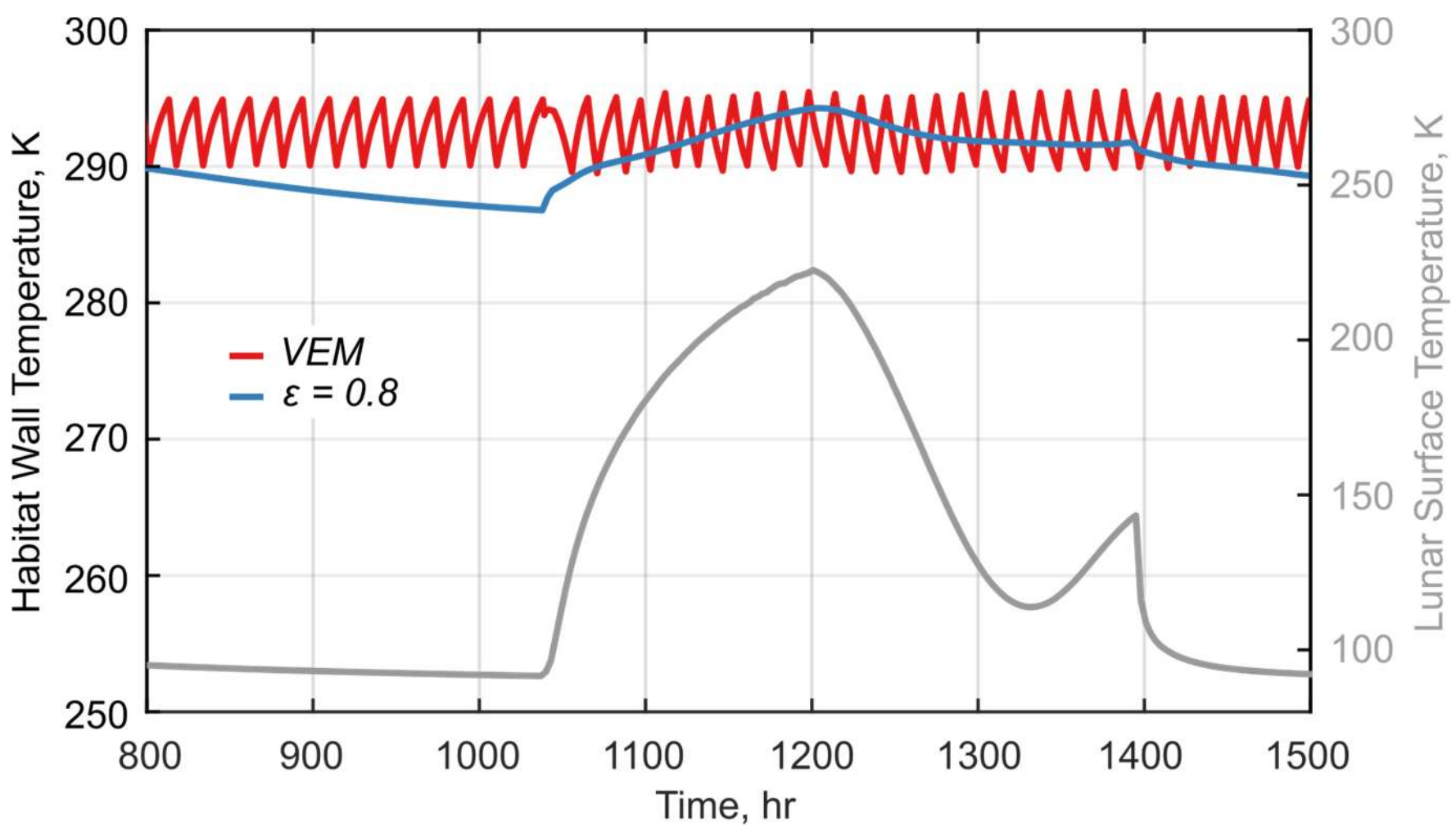

**Figure 8. Habitat internal wall temperature over one lunar day-night cycle using constant high-emissivity radiators and variable emissivity radiators. Lunar surface temperature is plotted on the right axis to indicate daytime.**

## IV. Conclusion

Throughout this work, we have described how to model variable emissivity materials using software such as *COMSOL Multiphysics®* and *Ansys Thermal Desktop®*. In particular, we investigate how VEMS could be used on lunar habitats for night survival and reduced power consumption. A topographic lunar surface model is used to model the surface temperatures of a real location near the lunar south pole. We then match the slope and shading of a planar model to the transient surface temperature of a specific region within the topographic study. This procedure results in a simplified thermal model of the lunar surface, which accounts for view factors resulting from regional features such as hills and craters. Using representative thermal properties, we develop a simplified dome-shaped habitat loosely based on previous conceptual designs. By combining this habitat with the simplified lunar surface model, we provide a platform to investigate variable emissivity materials and other novel radiator coatings.

We then present a generalized framework for modeling variable emissivity materials in *COMSOL Multiphysics®* and *Ansys Thermal Desktop®*. The *Events* interface and *Dynamic Sinda* are used in each respective program to pause the transient calculation once the emissivity switching criteria are met. The thermal emissivity is updated, and radiosity is calculated until the next time step at which the emissivity is expected to switch. We note that the variable emissivity simulations performed throughout this study were conducted on different systems for CM and TD, respectively, and as such, a direct comparison of computation speed and cost cannot be presented. The primary goal of this work was to demonstrate that both programs can be used in future variable emissivity studies, and our results are independent of the specific software used. As habitat designs become increasingly complex and requires various integrated systems, it is expected that TD will become more practical; however, for high-level conceptual applications, both CM and TD may be used. We used both software to conduct two case studies controlling VEM properties with respect to time and temperature to show their benefit for sustainable lunar habitats and other long-term extraterrestrial systems. We show that switching to a low thermal emissivity state for the duration of lunar night has the potential to reduce outgoing heat flux by ~1.36 kW/m$^2$ when the habitat interior temperature is prescribed as 300 K. Lastly, we present the ability to switch emissivity state in response to temperature. This approach allows for more dynamic control over emissivity state, resulting in the ability to achieve near-constant temperatures over a full lunar day-night cycle.

The methods and results presented in this study aim to increase the Technology Readiness Level (TRL) of VEMs for lunar applications and further their adoption across new technologies designed for space. We focus on demonstrating the potential to assist with lunar night survival; however, by modifying the environment, we envision this approach could be used to demonstrate the benefit of using VEMs for thermal management on Mars. In general, the steps described serve as general guidelines for modeling VEMs on extraterrestrial bodies, and we envision they could be expanded to include more realistic habitats, new locations on the moon, or temperature-dependent emissivity of specific radiator materials. Despite the clear benefit of variable emissivity materials for lunar night survival, there remain several key practical aspects limiting their use, which range from fabrication complexity and operating temperature to lunar dust removal. In particular, fabricating variable emissivity materials, such as those comprising Vanadium Dioxide, at scalable quantities has not yet been demonstrated, and while doping may reduce the phase transition temperature, it also decreases the emissivity contrast between states. Additionally, lunar dust buildup on externally facing radiators is expected to significantly impact performance. With strong solar absorptivity, we envision that significant dust coverage could lead to undesirable daytime heating. To overcome this limitation, we expect variable emissivity radiators will be combined with novel dust mitigation strategies such as Electrodynamic Dust Shields (EDS) or various forms of mechanical removal.

## Acknowledgments

SK acknowledges support through the NASA Space Technology Graduate Research Opportunities (NSTGRO) fellowship 80NSSC24K1372. The authors acknowledge support by DARPA (award number D24AP00310-00) and the Office of Naval Research (award number N00014-23-1-2627).